\documentclass[submission,letterpaper,Phys]{SciPost}
\pdfoutput=1
\usepackage{amsmath,amssymb,mathtools,xspace}
\usepackage{amsthm}
\usepackage{booktabs,multirow,graphicx,tabularx,slashed}
\usepackage{hyperref}
\usepackage{color,xcolor}
\usepackage[normalem]{ulem}
\usepackage{enumitem}
\usepackage{feynmp}
\usepackage{braket}
\usepackage{tikz}
\usepackage{booktabs}
\usepackage[colorinlistoftodos]{todonotes}

\usetikzlibrary{shapes.geometric, arrows}

\usepackage{algorithm}
\usepackage{algorithmic}

\def\tm{\tau_{min}\xspace}

\makeatletter
\@ifundefined{pdfoutput}{}{\DeclareGraphicsRule{*}{mps}{*}{}}
\makeatother

\definecolor{Gcolor}{HTML}{3b528b}
\definecolor{Dcolor}{HTML}{e41a1c}

\setitemize{itemsep=2pt,topsep=2pt,parsep=0pt,partopsep=0pt,labelindent=2pt,leftmargin=*}
\setenumerate{itemsep=0pt,topsep=2pt,parsep=0pt,partopsep=0pt,labelindent=3pt,leftmargin=*}

\tikzstyle{generator} = [rectangle, rounded corners, minimum width=3cm, minimum height=1cm,text centered, draw=Gcolor]
\tikzstyle{discriminator} = [rectangle, rounded corners, minimum width=3cm, minimum height=1cm,text centered, draw=Dcolor]
\tikzstyle{io} = [circle, trapezium left angle=70, trapezium right angle=110, minimum width=1cm, minimum height=1cm, text centered, draw=black]

\tikzstyle{process} = [rectangle, minimum width=1cm, minimum height=1cm, text centered, draw=black]
\tikzstyle{decision} = [rectangle, minimum width=1cm, minimum height=1cm, text centered, draw=black]

\tikzstyle{arrow} = [thick,->,>=stealth]
\usepackage{xcolor}

\newcommand{\Langle}{\big\langle}
\newcommand{\Rangle}{\big\rangle}

\newcommand\one{\leavevmode\hbox{\small1\normalsize\kern-.33em1}}

\newcommand{\qqquad}{\qquad \qquad}

\newcommand{\gev}{\text{GeV}}

\def\slashchar#1{\setbox0=\hbox{$#1$}           
   \dimen0=\wd0                                 
   \setbox1=\hbox{/} \dimen1=\wd1               
   \ifdim\dimen0>\dimen1                        
      \rlap{\hbox to \dimen0{\hfil/\hfil}}      
      #1                                        
   \else                                        
      \rlap{\hbox to \dimen1{\hfil$#1$\hfil}}   
      /                                         
   \fi}

\newcommand{\ie}{\textsl{i.e.}\;}

\renewcommand{\d}{{\text{d}}}

\newcommand{\vegas}{\texttt{Vegas}\xspace}
\newcommand{\lhapdf}{\texttt{LHAPDF}\xspace}

\newcommand{\sherpa}{\texttt{Sherpa}\xspace}

\begin{document}

\begin{center}{\Large \textbf{
How to GAN Event Unweighting
}}\end{center}

\begin{center}
Mathias Backes\textsuperscript{1},
Anja Butter\textsuperscript{1},
Tilman Plehn\textsuperscript{1}, and
Ramon Winterhalder\textsuperscript{1}
\end{center}

\begin{center}
{\bf 1} Institut f\"ur Theoretische Physik, Universit\"at Heidelberg, Germany
winterhalder@thphys.uni-heidelberg.de
\end{center}

\begin{center}
\today
\end{center}


\section*{Abstract}
         {\bf Event generation with neural networks has seen
           significant progress recently. The big open question is
           still how such new methods will accelerate LHC simulations
           to the level required by upcoming LHC runs. We target a
           known bottleneck of standard simulations and show how their
           unweighting procedure can be improved by generative
           networks. This can, potentially, lead to a very significant
           gain in simulation speed.}

\vspace{10pt}
\noindent\rule{\textwidth}{1pt}
\tableofcontents\thispagestyle{fancy}
\noindent\rule{\textwidth}{1pt}

\newpage
\section{Introduction}
\label{sec:intro}

First-principle simulations have defined data analysis at the LHC
since its beginning. The success of the LHC in establishing the
Standard Model as the fundamental theory of particle interactions is
largely owed to such precision simulations and the qualitative
progress in our understanding of QCD. Because the HL-LHC will produce
a data set more than 25 times the current Run~2 data set, the current
theory challenge is to provide significantly faster simulations, while
at the same time increasing the precision to the per-cent level and
better. This goal is defined by QCD precision predictions as well as
by the expected size of experimental uncertainties, which seriously
limit the use of leading-order simulations even for complex signatures
at future LHC runs. While it is hard to accelerate standard tools to
the required level, there is justified hope that modern machine
learning will allow us to reach this goal.

A range of modern neural network applications to LHC simulations have
been proposed over the last two years~\cite{Butter:2020tvl}. The
conceptually most ambitious network architecture are generative
networks, like generative adversarial networks
(GAN)~\cite{Goodfellow:2014:GAN:2969033.2969125,Creswell2018,Butter:2020qhk},
variational autoencoders~\cite{kingma2014autoencoding,Kingma2019}, or
normalizing flows~\cite{10.5555/3045118.3045281,Kobyzev_2020,nflow1}
including invertible neural networks (INNs)~\cite{inn,coupling2,glow}.
Following the established Monte Carlo structures leads us to consider
phase space integration~\cite{maxim,bendavid}, phase space
sampling~\cite{Bothmann:2020ywa,Gao:2020vdv,Gao:2020zvv,Chen:2020nfb},
and amplitude networks~\cite{Bishara:2019iwh,Badger:2020uow}. A
technical step beyond the standard approach are fully network-based
event
generation~\cite{dutch,gan_datasets,DijetGAN2,Butter:2019cae,Alanazi:2020klf,Kansal:2020svm},
including event subtraction~\cite{subgan}, detector
simulation~\cite{calogan1,calogan2,fast_accurate,aachen_wgan1,aachen_wgan2,ATLASShowerGAN,ATLASsimGAN,Belayneh:2019vyx,Buhmann:2020pmy,Diefenbacher:2020rna},
or parton
showering~\cite{shower,locationGAN,monkshower,juniprshower,Dohi:2020eda}.
Generative networks can also help to extract new physics
signals~\cite{bsm_gan} or experimental
anomalies~\cite{Nachman:2020lpy,Knapp:2020dde}.  Conceptually more
far-reaching, conditional generative networks can invert the forward
simulation chain to unfold detector effects~\cite{Datta:2018mwd,fcgan}
and extract the hard scattering at parton
level~\cite{Bellagente:2020piv}. Many of these applications are
currently finding their way into the LHC theory toolbox.

Going back to the LHC motivation, the key question is where we can
gain significant speed in precision theory simulations. As mentioned
above, we can use flow networks to improve the phase space
sampling~\cite{Bothmann:2020ywa,Gao:2020zvv}. In addition, we can
employ generative networks because they learn more information than a
statistically limited training data set~\cite{Butter:2020qhk}. This is
why neural networks are successfully used to encode parton
densities~\cite{DelDebbio:2004xtd}.  Finally, there exist promising
hints for network extrapolation in jet kinematics~\cite{DijetGAN2}.

In this paper we follow a different path and target a well-known
bottleneck LHC event simulation, the transformation of weighted into
unweighted events~\cite{Hoeche:2019rti, reweighting_ben}. Usually, the
information about a differential scattering rate is first encoded in a
combination of event weights and the event distribution over phase
space. To compare with data we ideally work with unit-weight events,
where all information is encoded in the event distribution. For
complex processes, the standard unweighting procedures suffer from low
efficiency, which means they lose statistical power. We will show how
a generative network, specifically a GAN, can unweight events without
these losses and thereby speed up event generation
significantly~\cite{matthias_thesis}. We will start with a
1-dimensional and a 2-dimensional toy model in
Sec.~\ref{sec:unweight}, to illustrate our uwGAN idea in the context
of standard approaches. In Sec.~\ref{sec:lhc_events} we will then use
a simple LHC application to show how our GAN-unweighting\footnote{in
  short GANweighting or GUNweighting.}  method can be applied to LHC
simulations.

\section{Unweighting GAN}
\label{sec:unweight}

Before we show how networks can be useful for LHC simulations, we
briefly introduce event unweighting as it is usually done, how a
generative network can be used for this purpose, and when such a
network can beat standard approaches. First, we will use a
1-dimensional camel distribution to illustrate the loss function which
is needed to capture the event weights. Second, we use a
2-dimensional Gaussian ring as a simple example where our method
circumvents known challenges of standard tools.

\subsection{Unweighting}
\label{sec:unweight_basics}

For illustration purpose, we consider an integrated cross section
of the form
\begin{align}
  \sigma
  = \int d x\,\frac{d \sigma}{d x}
  \equiv \int d x\, w(x) \; ,
\label{eq:event_generation}
\end{align}
where $d\sigma/d x$ is the differential cross section over the
$m$-dimensional phase space $x$. To compute this integral numerically
we draw $N$ phase space points or events $\{x\}$ and evaluate
\begin{align}
  \sigma
  \approx \left\langle \frac{d\sigma}{d x} \right\rangle
  \equiv \langle w(x) \rangle \; .
\end{align}
The event weight $w(x)$ describes the probability for a single event
$x$. Sampling $N$ phase space points $\{x\}$ and evaluating their
weights $\left\{w\right\}$ defines $N$ weighted events
$\left\{x,w\right\}$. The information on the scattering process is
encoded in a combination of event weights and phase space
density. This can be useful for theory computations, but actual events
come with unit weights, so all information is encoded in their phase
space density alone.

We can easily transform $N$ weighted events $\left\{x, w \right\}$
into $M$ unweighted events $\left\{x\right\}$ using a hit-or-miss
algorithm, where in practice $M \ll N$. It re-scales the weight $w$
into a probability to keep or reject the event $x$,
\begin{align}
w_\text{rel} =\frac{w}{w_\text{max}} \; ,
\end{align}
and then uses a random number $R\in[0,1]$ such that the event is kept
if $w_\text{rel}> R$. The obvious shortcoming of this method is that
we lose a lot of events. For a given event sample the unweighting
efficiency is~\cite{Bothmann:2020ywa}
\begin{align}
\epsilon_\text{uw}=\frac{\langle w \rangle}{w_\text{max}}\;.
\label{eq:uw_efficiency}
\end{align}
If the differential cross section varies strongly, $\langle w
\rangle\ll w_\text{max}$, this efficiency is small and the LHC
simulation becomes CPU-intensive.

A standard method to improve the sampling and integration are phase
space mappings, or coordinate transformations $x \to y(x)$,
\begin{align}
  \sigma
  = \int dx\,w(x)
  = \int dy\,\left|\frac{\partial x}{\partial y}\right|\,w(y)
  \equiv \int dy \; \tilde{w}(y) \;.
\end{align}
Ideally, the new integrand $\tilde{w}(y)$ is nearly constant and the
structures in $w(x)$ are fully absorbed by the Jacobian. In that case
\begin{align}
  \tilde\epsilon_\text{uw}
  =\frac{\langle\tilde w\rangle}{\tilde{w}_\text{max}}
  \approx \frac{\langle C\rangle}{C} =1 \; .
\end{align}
This method of choosing an adequate coordinate transformation is
called importance sampling. The most frequently used algorithm is
\vegas~\cite{vegas1,vegas2}, which assumes that $g(x)$ factorizes into
phase space directions, as we will discuss later.\medskip

In contrast to vetoing most of the weighted events we propose to use
all of them to train a generative model to produce unweighted events.
We follow a standard GAN setup with spectral normalization~\cite{spectralnorm}
as regularization method
\begin{align}
\begin{split}
L_D &=   \Langle -\log D(x) \Rangle_{x \sim P_T} 
+ \Langle - \log (1-D(x)) \Rangle_{x \sim P_G}\;, \\
L_G &= \Langle - \log D(x) \Rangle_{x \sim P_G} \; .
\label{eq:GAN_normal}
\end{split}
\end{align}
For weighted training events, the information in the true distribution $P_T$
factorizes into the distribution of sampled events $Q_T$ and their
weights $w(x)$. To capture this combined information we replace the
expectation values by weighted means for batches of weighted events,
\begin{align}
\begin{split}
L_D^{(\text{uw})}  &=  \frac{\langle -w(x)\,\log D(x)\rangle_{x\sim Q_T}}{\langle w(x)\rangle_{x\sim Q_T}}
+ \Langle -\,\log (1-D(x)) \Rangle_{x \sim P_G}\;,\\
L_G^{(\text{uw})} &= \Langle - \log D(x) \Rangle_{x \sim P_G} \; .
\label{eq:uwGAN}
\end{split}
\end{align}
Because the generator produces unweighted events with $w_G(x)=1$ their weighted
mean reduces to the standard expectation value. This way, our
unweighting GAN (uwGAN) unweights events, and the standard GAN is just
a special case with all information encoded in the event distribution.

\subsection{One-dimensional camel back}
\label{sec:unweight_1d}

\begin{figure}[t]
\includegraphics[page=1, width=.32\textwidth]{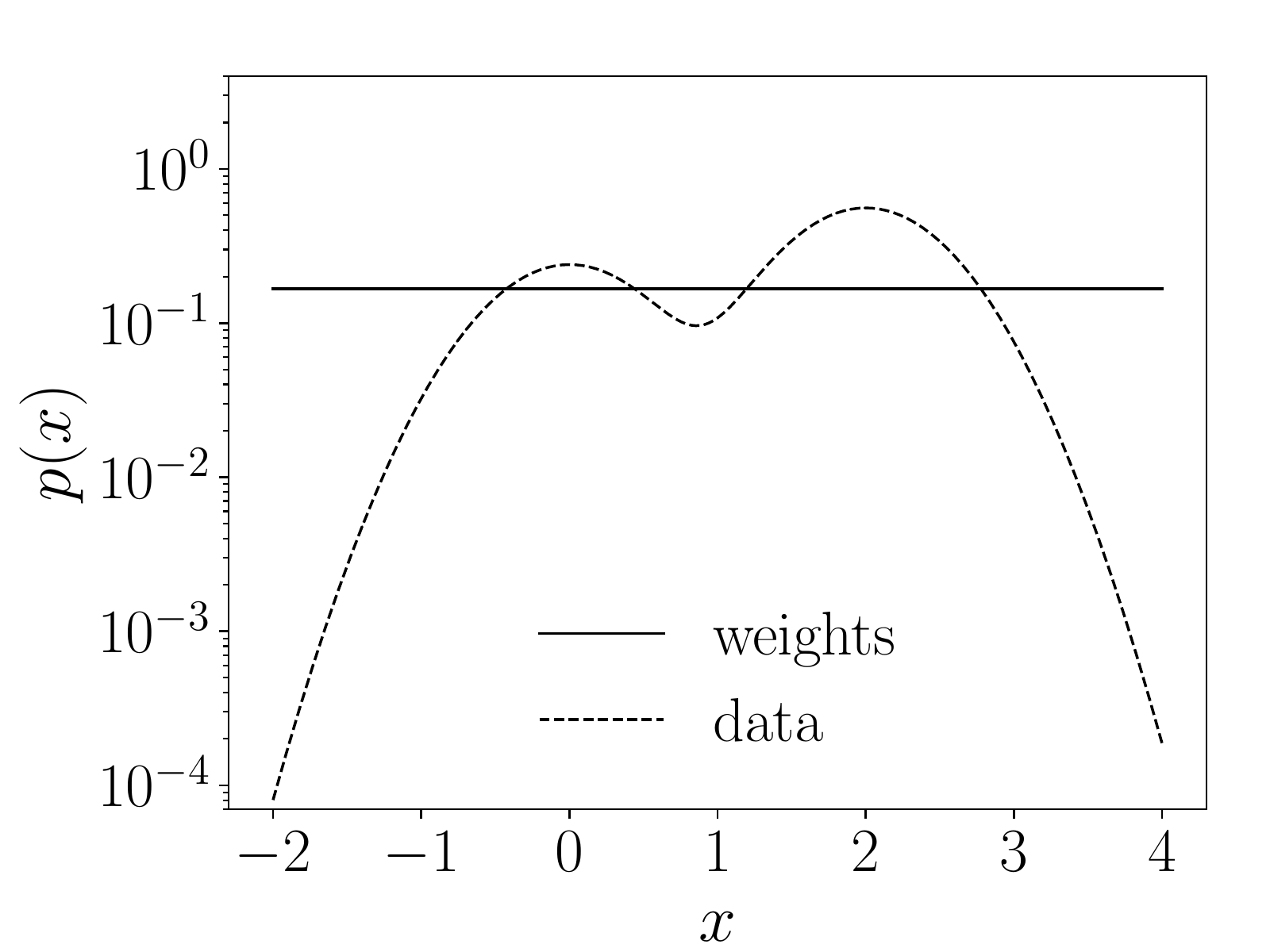}
\includegraphics[page=1, width=.32\textwidth]{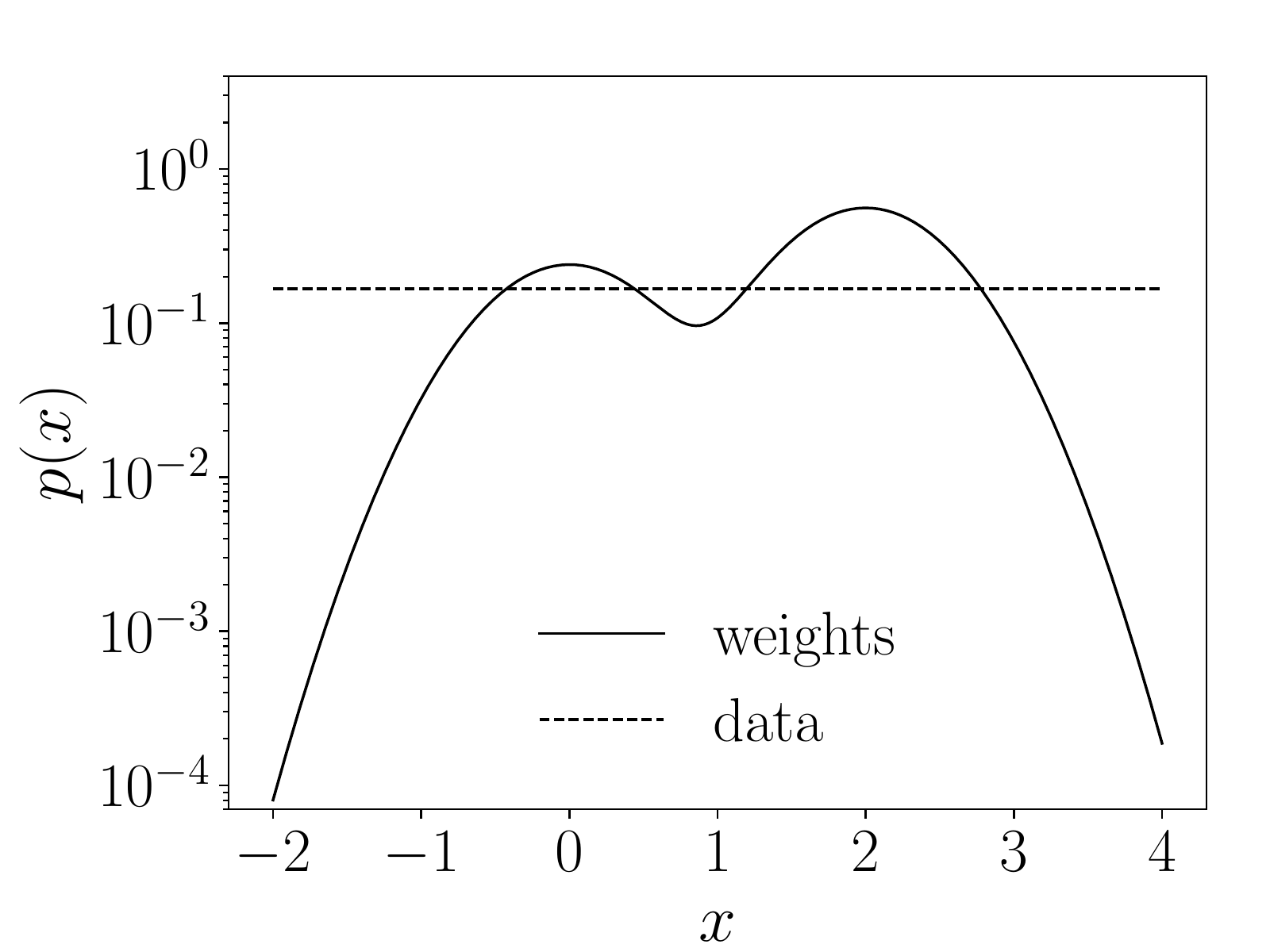}
\includegraphics[page=1, width=.32\textwidth]{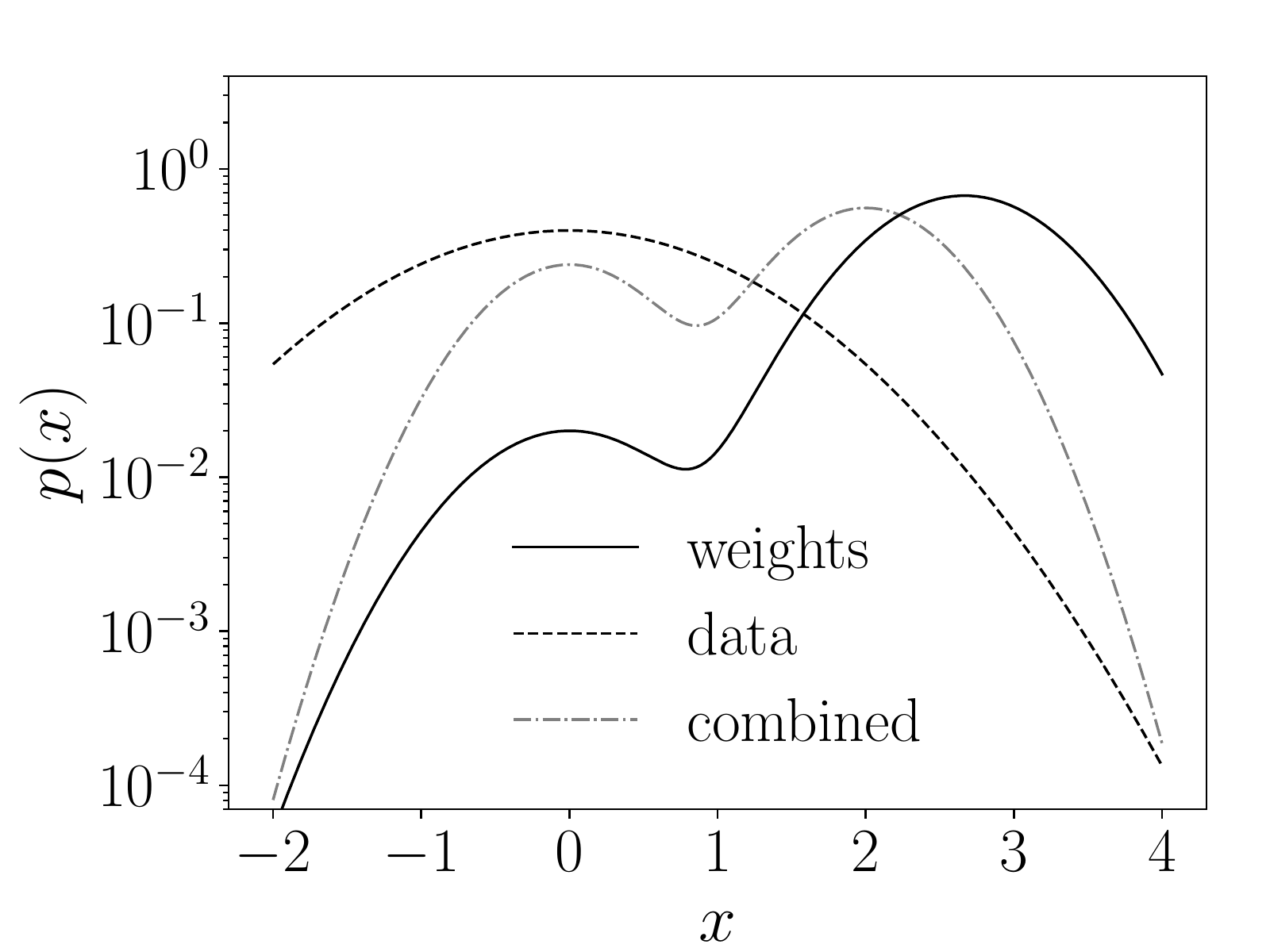}\\
\includegraphics[page=1, width=.32\textwidth]{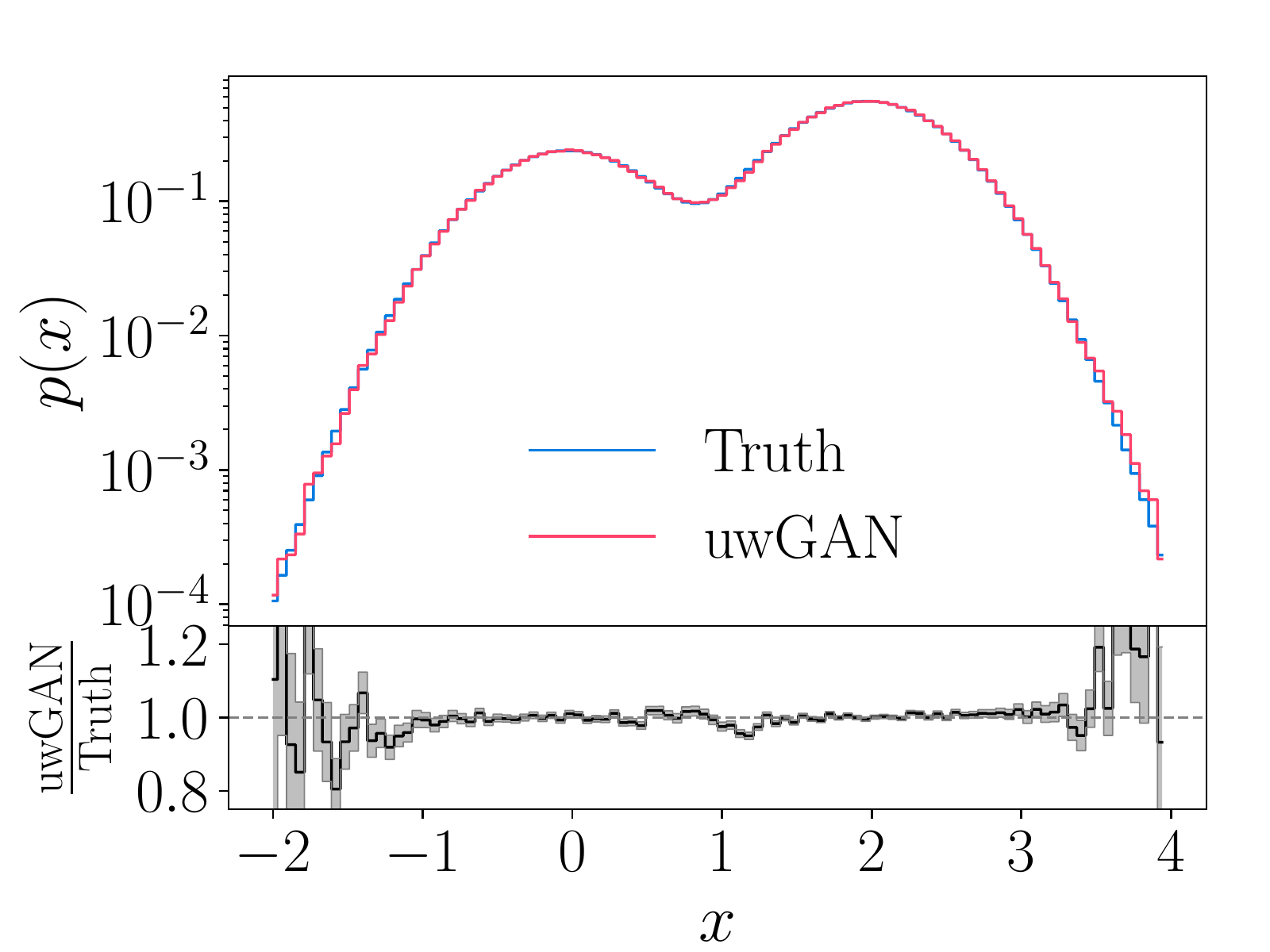}
\includegraphics[page=1, width=.32\textwidth]{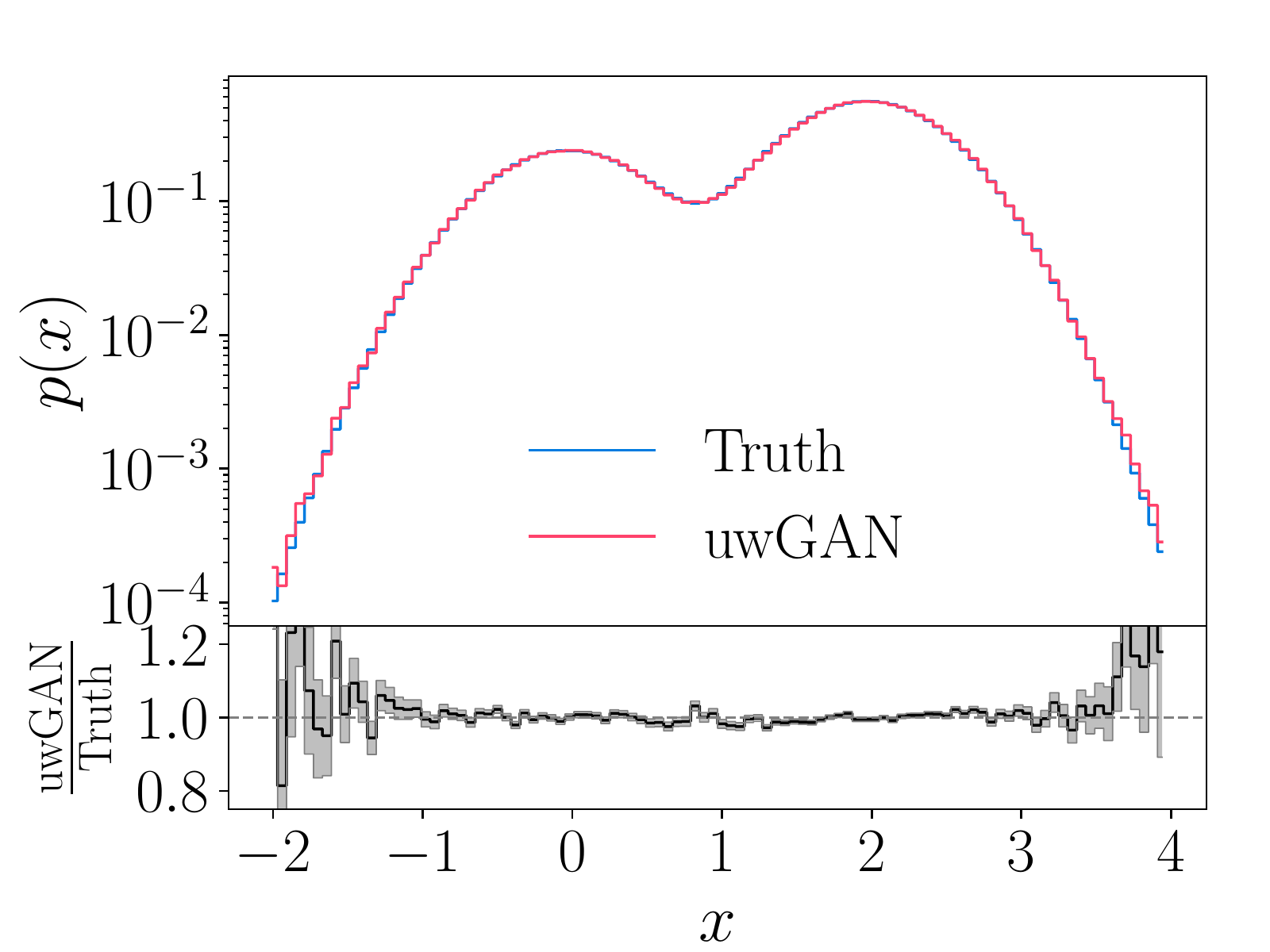}
\includegraphics[page=1, width=.32\textwidth]{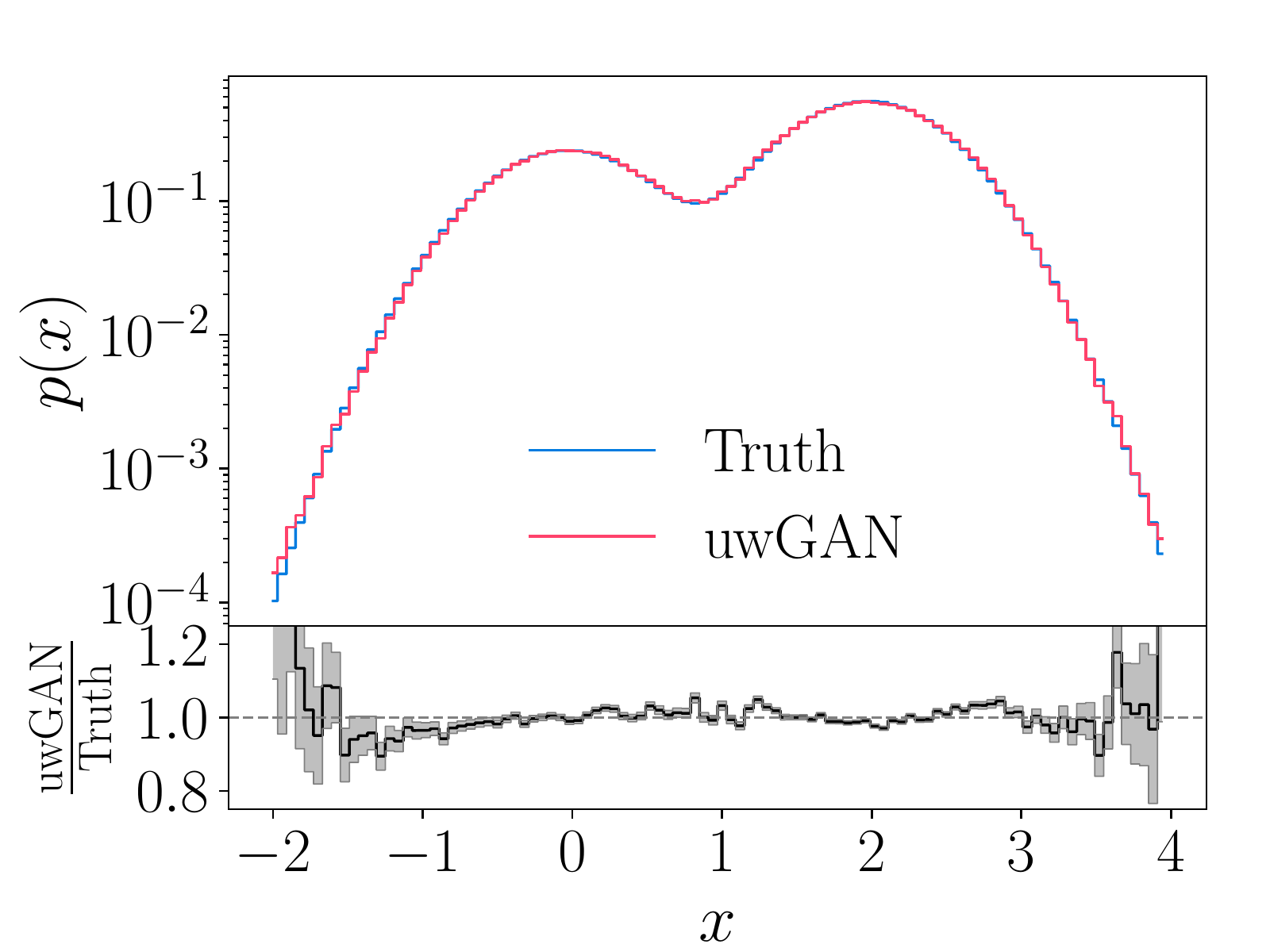}
\caption{Event vs weight distribution of the training data (top) and GANned vs truth kinematic
  distributions (bottom) for unweighted events (left), uniform
  distribution (center), and our hybrid case (right). The lower panels
  in the bottom show the bin-wise truth-to-GAN ratio.}
\label{fig:camel}
\end{figure}

We illustrate the unweighting GAN with a 1-dimensional camel
back
\begin{align}
P_\text{camel}(x)=0.3\;\mathcal{N}(x;\mu=0,\sigma=0.5) + 0.7\; \mathcal{N}(x;\mu=2,\sigma=0.5),
\label{eq:uw_camel}
\end{align}
where $\mathcal{N}(x;\mu,\sigma)$ is a Gaussian.  To see how the GAN
reacts to different ways of spreading the information between weights
and distributions, we define three events samples,
\begin{itemize}
\item[$\cdot$] unweighted events distributed according to
the camel back, $X_\text{u}=(x_\text{camel},
w_\text{uniform})$;
\item[$\cdot$] uniformly distributed events
  $X_\text{w}=(x_\text{uniform}, w_\text{camel})$; and
\item[$\cdot$] a split
$X_\text{hybrid}=(x_{q_1}, w_{q_2})$ with $P_\text{camel}(x) \propto
  q_1(x) q_2(x)$ and $q_1(x) = \mathcal{N}(x;\mu=0,\sigma=1)$.
\end{itemize}
For the camel back example our training data will consist of 1 million
weighted events. We use 32 units within 2 layers in the generator and 32
units within 3 layers in the discriminator. In the hidden layers, we
employ the standard ReLU activation function, $\max(0,x)$, for both networks.  To compensate
an imbalance in the training we update the discriminator ten times as
often as the generator. As a first test, we show in
Fig.~\ref{fig:camel} how our GAN reproduces the full 1-dimensional
target distribution from unweighted events, uniformly distributed
events, and weighted events equally well. The limitation are always the
poorly populated tails in the training data.

\begin{figure}[t]
  \centering
  \includegraphics[width=.49\textwidth]{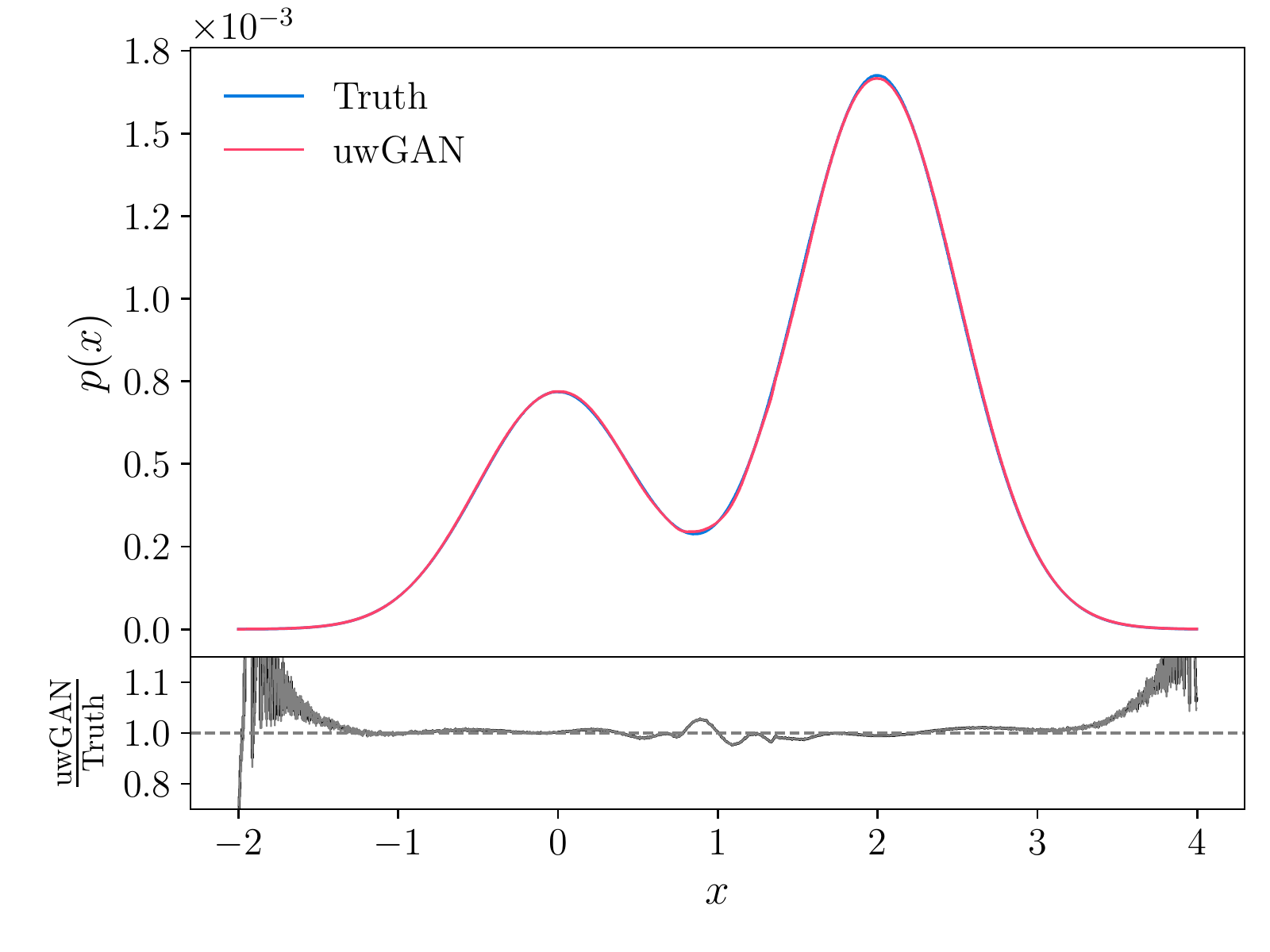}
  \includegraphics[width=.49\textwidth]{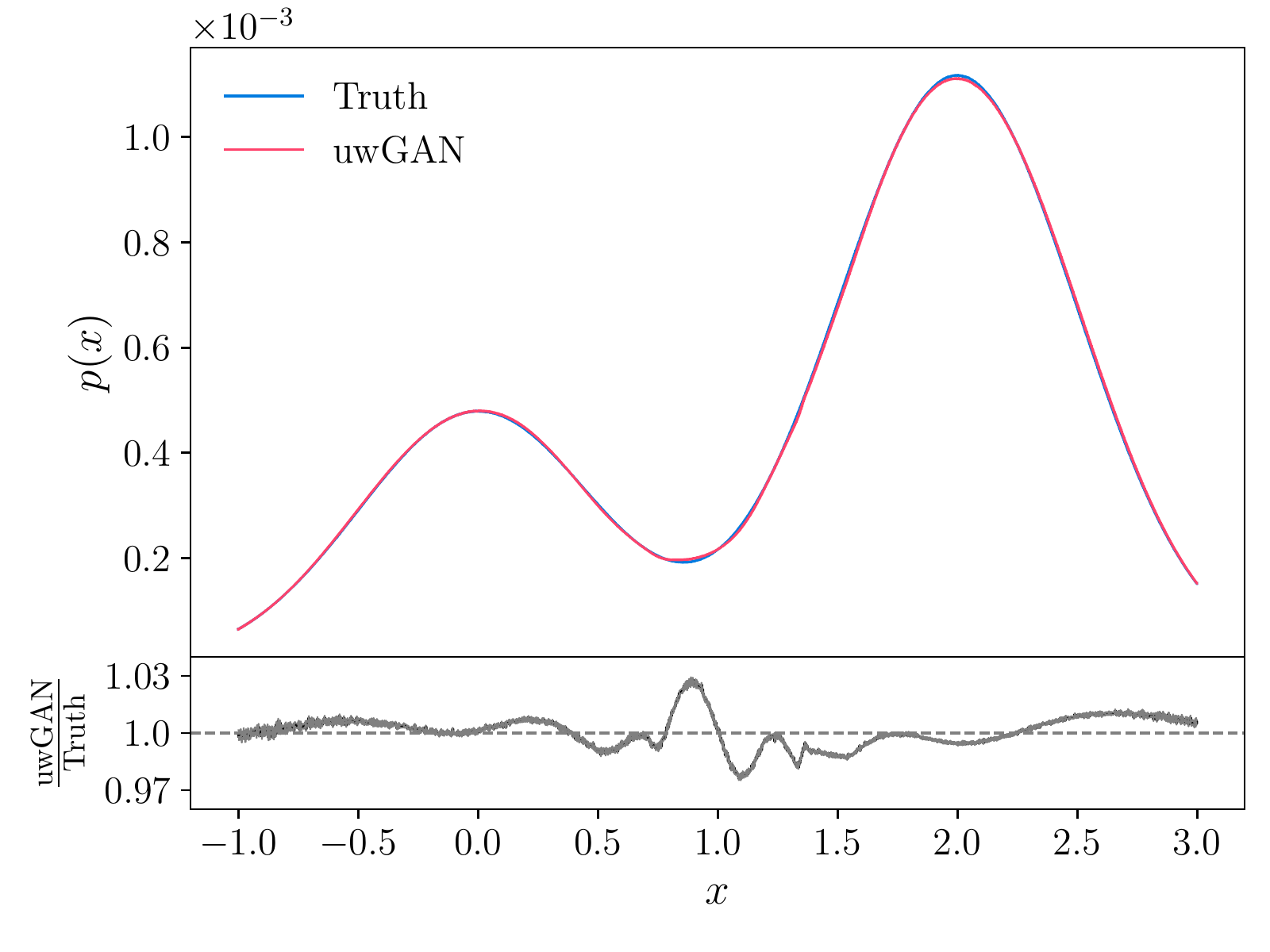}\\
  \includegraphics[width=.49\textwidth]{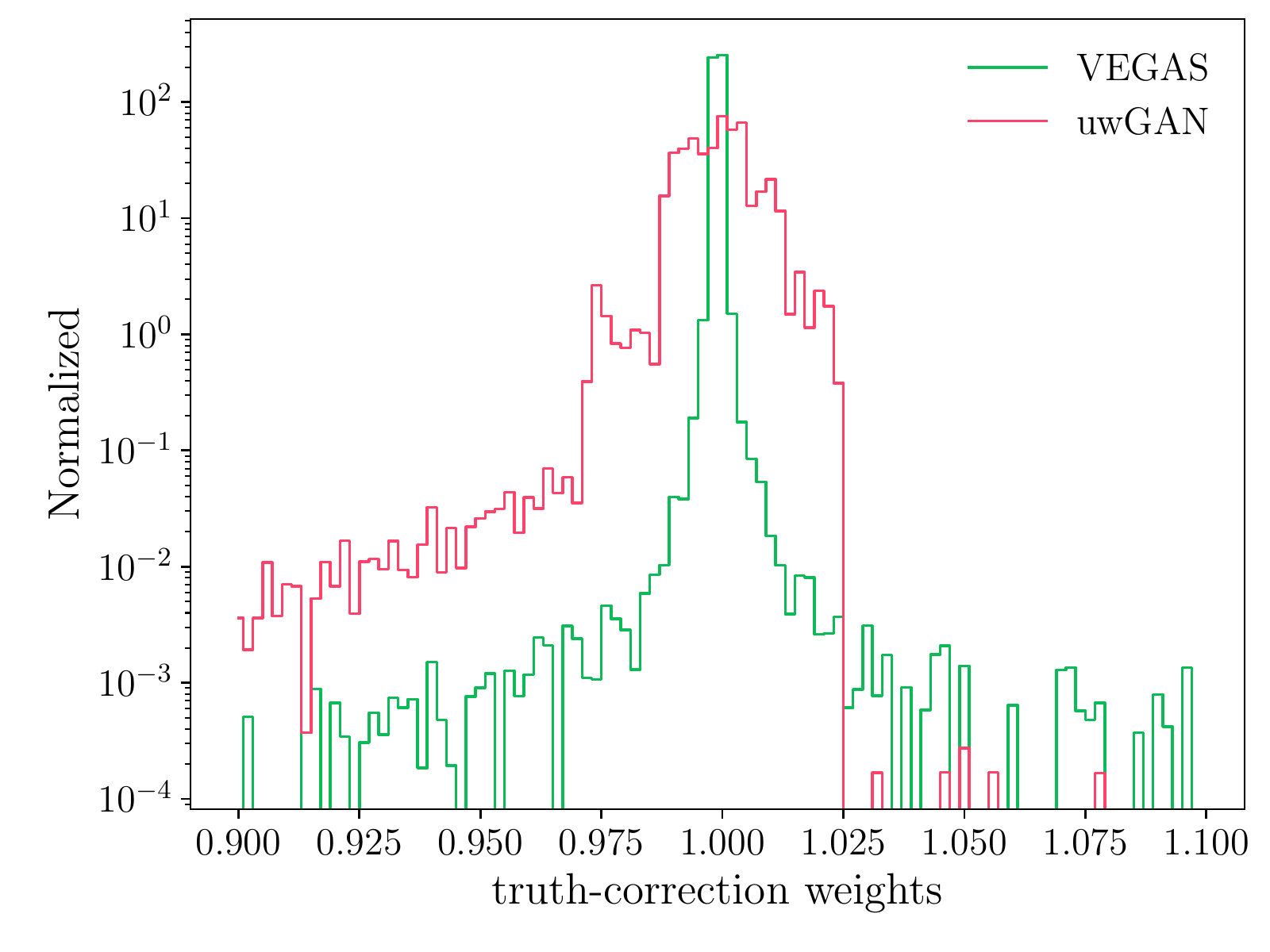} 
  \includegraphics[width=.49\textwidth]{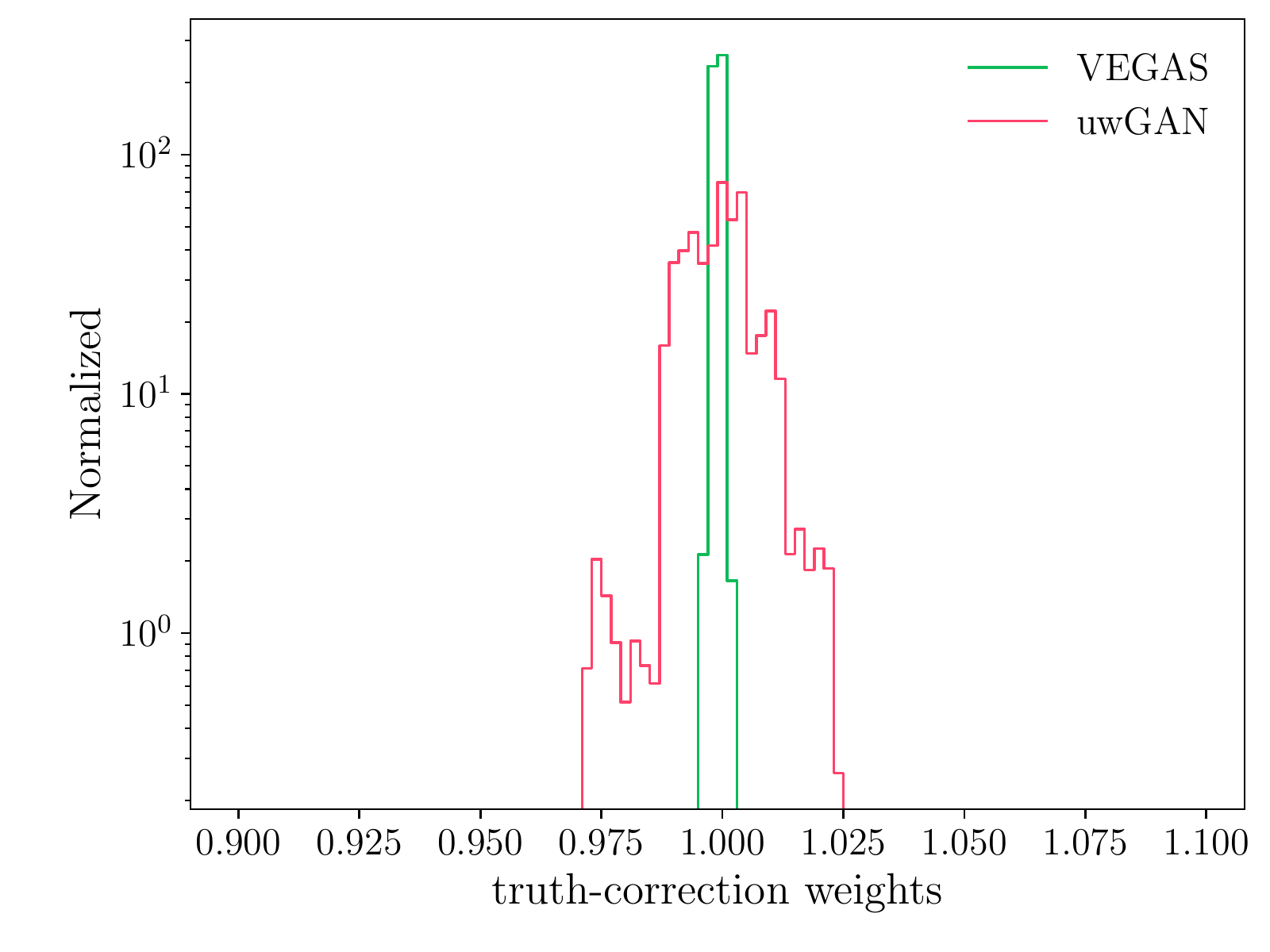}
  \caption{Upper: training and GANned distribution for the
    1-dimensional camel back. In the right panels we remove the
    tails of the distribution. Lower: truth-correction weights
    for the GANned events, compared with the \vegas weight
    distribution.}
  \label{fig:uw_vegas}
\end{figure}

To benchmark our unweighting GAN, we first sample the true
distribution with a large number of events and bin them finely, in our
1-dimensional case $10^{10}$ events in $2000$ bins equally distributed
over the full range $x=-2~...~4$. This statistics goes far beyond the
training sample and is only needed to define a truth benchmark. We
then generate an equally large sample of $10^{10}$ GAN events and
compare the two high-statistics samples, weighted truth events and
unweighted GANned events, in the top panels of
Fig.~\ref{fig:uw_vegas}. From the bin-wise ratio we see that the GAN
reproduces the true distribution at the few per-cent level, again
limited by the tails.

Given the true distribution, we can compute event-wise factors which
would be needed to shift the GANned unit weights to reproduce the
true distribution exactly. We refer to them as truth-correction
weights for each (unweighted) GAN event.  Because we rely on the
binned truth information we assign the same truth correction to all
GAN events in a given, narrow bin. Formally, we assume that the
generator distribution $P_G$ approximates the true distribution
$P_T$, so the bin-wise ratio
\begin{align}
w_G(x)=\frac{P_T(x)}{P_G(x)} 
\end{align}
for each unweighted event, given its phase space position $x$, should
tend to one. The actual values for the truth-correction weights are
shown in the bottom panels of Fig.~\ref{fig:uw_vegas}. For the full
$x$-range we see that they are strongly peaked around unity, but with
sizeable tails. The fact that the distribution is not symmetric and
includes significant statistical fluctuations suggests that our
network could be further improved. Nevertheless, the vast majority of
events have a truth-correction below 3\%. In the right panel we see
the same distribution after removing the tails. Literally all GAN
events now come with a truth-correction below 3\%. Comparing the
upper and lower panels of Fig.~\ref{fig:uw_vegas} we also see that
these truth-correction weights are not statistically distributed
corrections, fluctuating rapidly as a function of $x$. Instead, they
reflect systematic limitations to the precision with which the GAN
learns $P_T(x)$ and encodes it into the phase space distribution.

As discussed above, \vegas encodes $P_T(x)$ jointly into the phase
space distribution and event weights~\cite{vegas1,vegas2}. This means
we can compare the GAN and \vegas encodings in the phase space
distribution by comparing the truth-correction weights in the sense
that for \vegas they will define the perfectly trained output.  After
a series of 150 adaption steps, \vegas reaches the weight distribution
shown in Fig.~\ref{fig:uw_vegas}, corresponding to an unweighting
efficiency of $0.75$. Note that after 50 adaption steps, this \vegas
unweighting efficiency was $0.95$. The reason is that \vegas is
optimized for integration by using tight grids in the bulk and wide
grids in the tails. The longer \vegas adapts its grid, the more events
are removed from the tails. This improves the numerical integration at
the cost of the unweighting efficiency. Indeed, in
Fig.~\ref{fig:uw_vegas} we see that the high-weight tails of the
\vegas truth-correction are comparable to the GAN case.  Again the
tails in the event weights correspond directly to the tails of the
density distribution over $x$. When it comes to unweighting the \vegas
events, these tails become a major problem, because they drive the
denominator in Eq.\eqref{eq:uw_efficiency}.

\subsection{Two-dimensional Gaussian ring}
\label{sec:unweight_2d}

\begin{figure}[t]
\centering
\includegraphics[width= 0.49\linewidth]{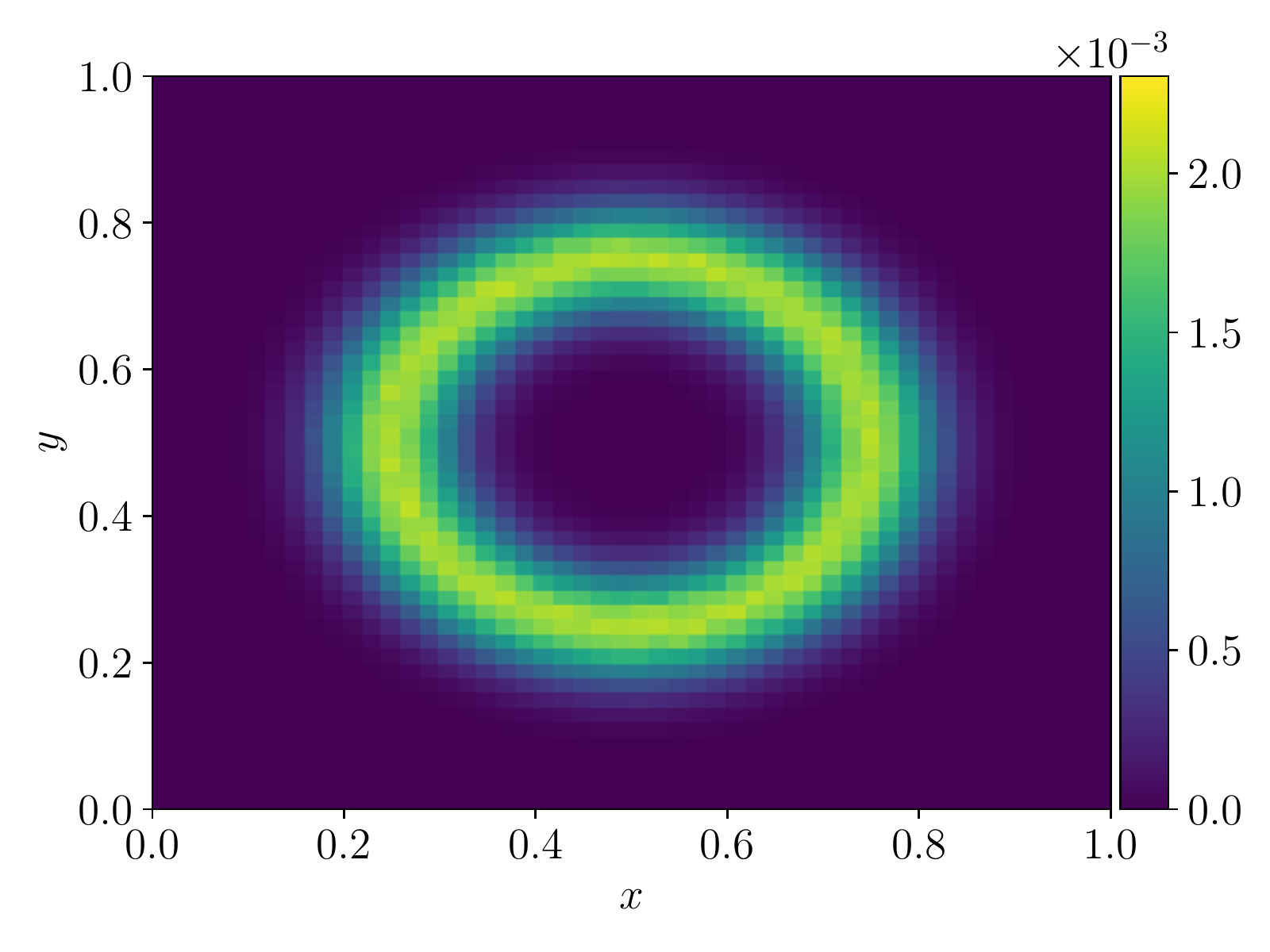}
\includegraphics[width= 0.49\linewidth]{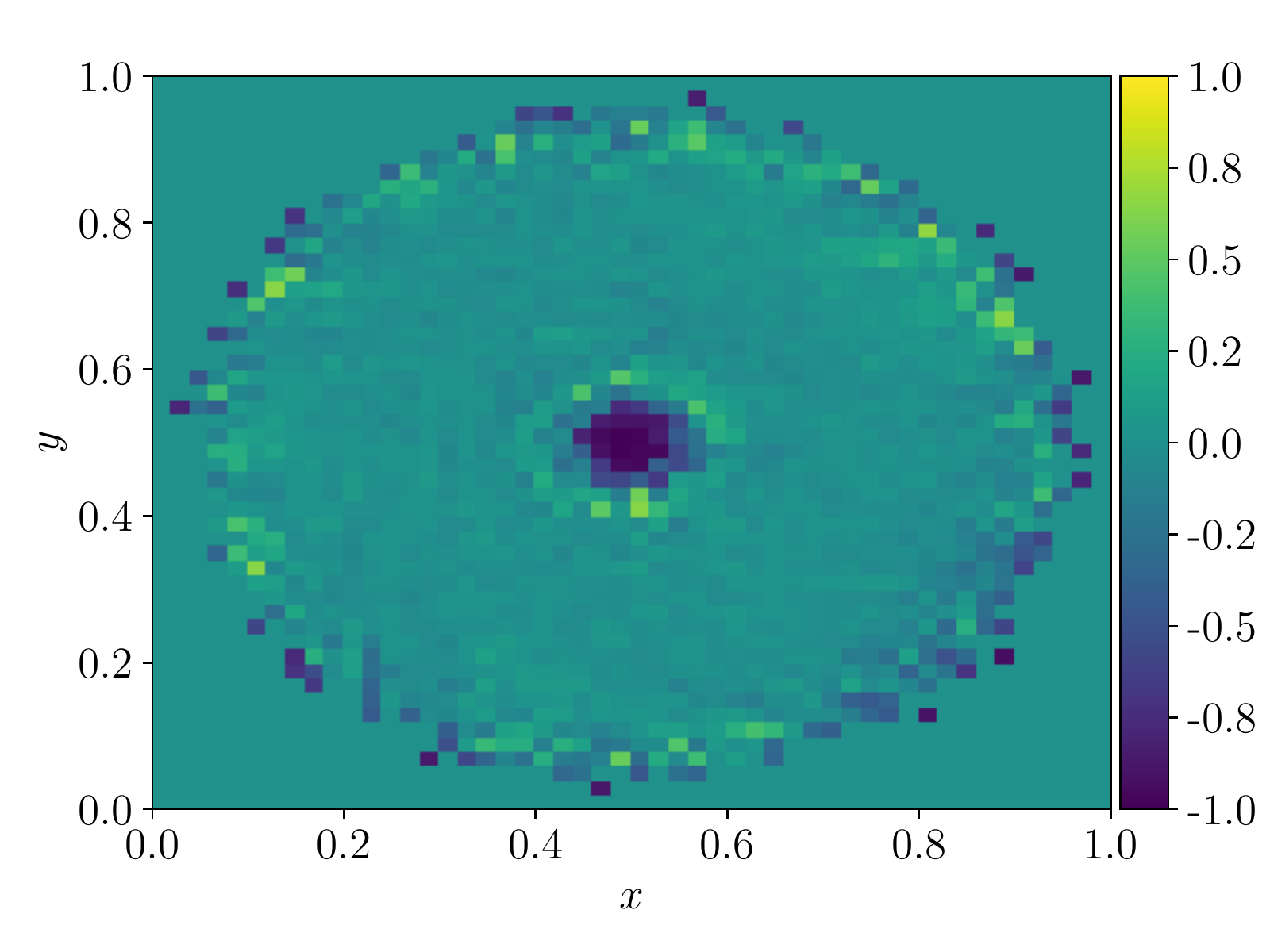} \\
\includegraphics[width= 0.49\linewidth]{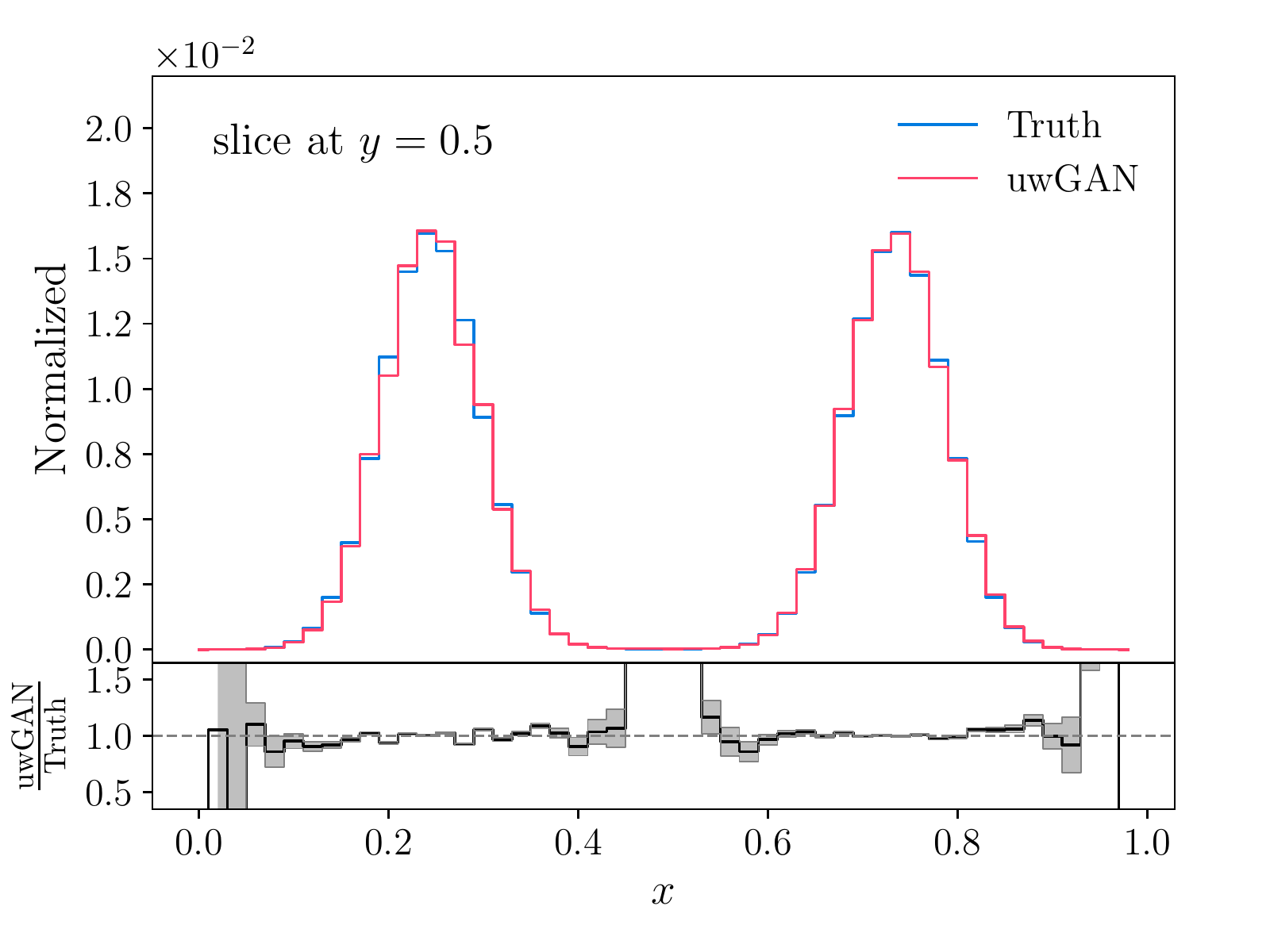}
\includegraphics[width=0.49\textwidth]{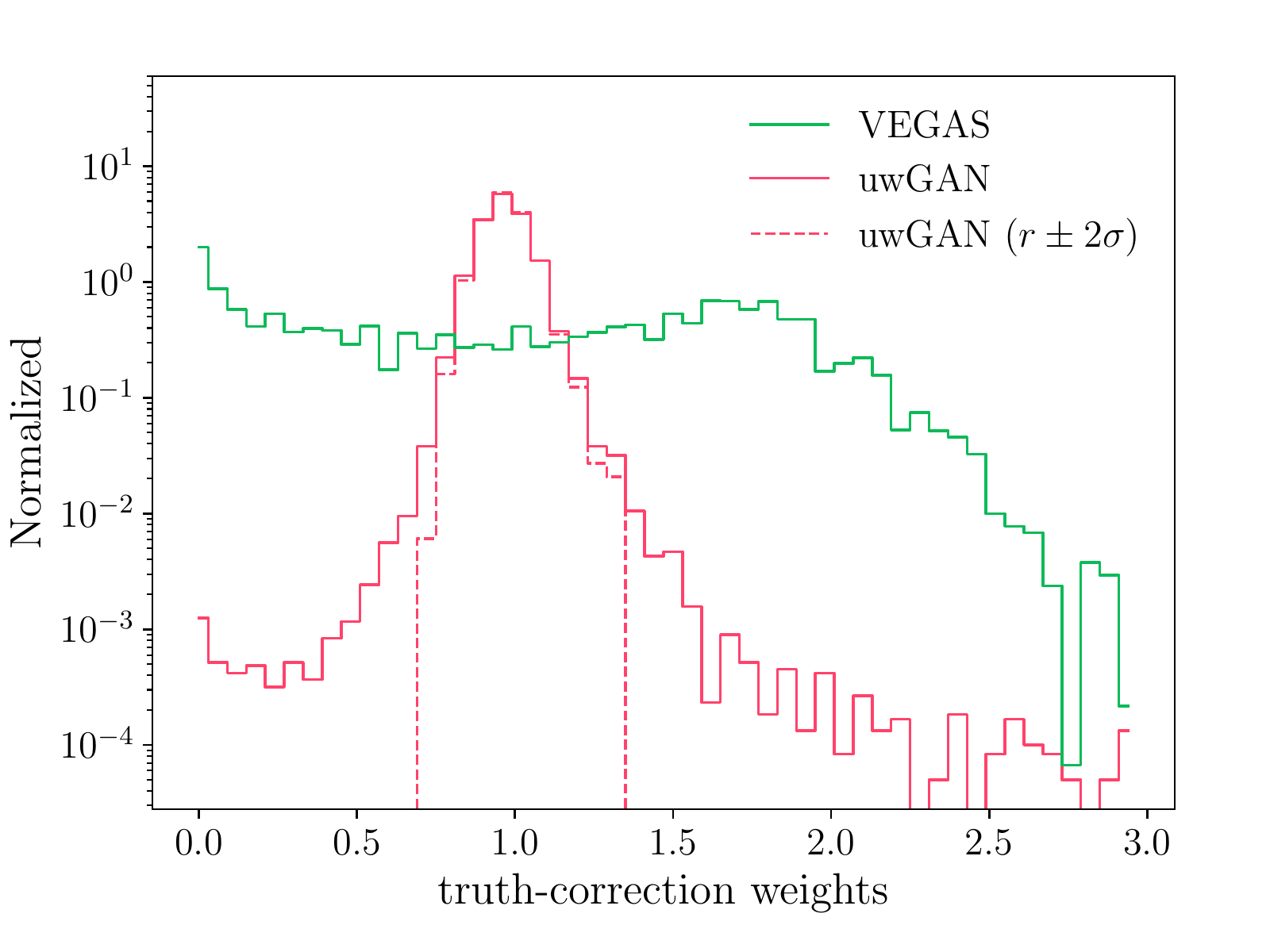}
\caption{Results for the 2-dimensional Gaussian ring showing the truth
  data (upper left), the asymmetry between truth and uwGAN (upper
  right), a 1-dimensional slice at $y=0.5$ (lower left), and the
  comparison of the truth-correction for uwGANned events with \vegas
  events (lower right). The dotted curve includes the bulk region
  $r \in [r_0 - 2\sigma, r_0 + 2\sigma]$ only.}
\label{fig:uw_gan_2D}
\end{figure}

Knowing a weakness of \vegas we now choose a 2-dimensional circle in the
$x$-$y$ plane with a Gaussian radial distribution as our second example,
\begin{align}
P_\text{circle}(x,y) = N \,  \exp \left[ -\frac{1}{2 \sigma^2} \, \left (\sqrt{(x-x_0)^2+(y-y_0)^2}-r_0 \right)^2 \right] \; ,
\label{eq:uw_circle}
\end{align}
with $x_0=y_0=0.5$, $r_0=0.25$, and $\sigma=0.05$. The normalization
is then given by $N \approx 5.079$. We use the same GAN architecture
as before, but with 256 units in 8 layers in, both, generator and
discriminator.

In Fig.~\ref{fig:uw_gan_2D} we show the true distribution as well as
the asymmetry of the truth and GANned distributions. As for the
1-dimensional camel back, large relative deviations are limited to the tail of
the distribution, in this case including the center of the circle. In
the lower-left panel of Fig.~\ref{fig:uw_gan_2D} we see how these
regions contribute little to the integral over the density.

It is clear that the \vegas algorithm cannot reproduce the circular
shape, because it breaks the factorization with the
dimensionality. Instead, \vegas constructs a square with a low
unweighting efficiency. Again, we compare the GAN and \vegas
truth-correction weights in the lower-right panel of
Fig.~\ref{fig:uw_gan_2D}. As expected, the uwGAN now does
significantly better, albeit with truth corrections up to $\pm 25\%$
in the tails. Just like for the 1-dimensional example, the tails in
the truth-correction correspond directly to the tails in the density,
so they reflect the statistical limitations of the training sample.
For a realistic application the key question becomes how this kind of
truth correction compares to the standard approaches and if it is
sufficient given the general statistical limitations in poorly
populated phase space regions.

As a side remark, it is of course possible to compute the truth
corrections without binning for the 1-dimensional and 2-dimensional
toy models. However, for a realistic LHC problem that will in general
not be the case, so we stick to the binned definition throughout this
paper. We have explicitly tested that our binned distributions agree
with the exact truth-correction distributions for the two toy models.

\section{Unweighting Drell--Yan}
\label{sec:lhc_events}

So far, we have considered two toy examples to motivate our
uwGAN. Next, we need to apply it to a simple LHC process, where we can
study the phase space patterns in some detail.  We consider the
Drell--Yan process
\begin{align}
p p  \to \mu^+ \mu^- \; .
\end{align}
We generate 500k weighted events at a CM energy of 14~TeV. The 4-dimensional fiducial phase space is
defined by the minimal acceptance cut
\begin{align}
  m_{\mu\mu} > 50~\gev \; 
\label{eq:cuts}
\end{align}
to avoid the photon pole in the numerical event generation.
The technical requirement on the weighted training events is that they
should cover a wide range of weights, so we can test if the uwGAN can
deal with this practical challenge. This means we cannot use a
standard Monte Carlo, where sophisticated phase space mappings encode
$p_T$ and $m_{\mu\mu}$ very well.

We implement our own custom event generator in Python, extracting the
matrix elements from \sherpa~\cite{Bothmann:2019yzt}, the parton
densities from \lhapdf~\cite{lhapdf}, and employing the Rambo-on-diet
sampling~\cite{Rambo, RamboDiet}.  The integration over the parton
momentum fractions is symmetrized in terms of $\tau = x_1 x_2$ as the
first phase-space variable,
\begin{align}
  \sigma
  = \int\limits_{\tau_\text{min}}^1\frac{d \tau}{\tau}  \int\limits_{\tau}^1
  \frac{d x_1}{x_1} \;
  \sum_{a,b} x_1 f_a(x_1) \; x_2 f_b(x_2) \; \hat{\sigma}_{ab}(x_1 x_2 s) \; ,
\end{align}
with Eq.\eqref{eq:cuts} translating into $\tau_\text{min}\approx
0.00128$. Mapping the phase space onto a unit hyper-cube defines two
random numbers $r_{1,2}$ through
\begin{align}
  \tau=\tau_\text{min}^{r_1}
  \qqquad
  x_1 =\tau^{r_2} 
  \qqquad 
  x_2 =\tau^{1-r_2} \; ,
\end{align}
such that
\begin{align}
  \sigma = 2 \log \tau_\text{min}
  \int\limits_0^1d r_1\,r_1\int\limits_0^1d r_2 \; 
    \sum_{a,b} x_1 f_a(x_1) \; x_2 f_b(x_2) \; \hat{\sigma}_{ab}(x_1 x_2 s) \; .
    \label{eq:phase-space-param}
\end{align}
With an additional random number $r_3 =(\cos\theta+1)/2 $ we can parametrize the 4-dimensional phase space as
\begin{align}
p_T &= 2 E_\text{beam} \; \tau_\text{min}^{r_1/2} \sqrt{r_3 (1-r_3)} \notag \\
p_{z_1} &= E_\text{beam} \; \left( \tau_\text{min}^{r_1 r_2} r_3 + \tau_\text{min}^{r_1 (1-r_2)} (r_3-1) \right) \notag \\
p_{z_2}&= E_\text{beam} \; \left( \tau_\text{min}^{r_1 r_2} (1-r_3) - \tau_\text{min}^{r_1 (1-r_2)} r_3 \right) \notag \\
\phi &= 2 \pi r_4 \; .
\label{eq:event-param}
\end{align}
%

\begin{figure}[t]
  \includegraphics[page=1, width=.49\textwidth]{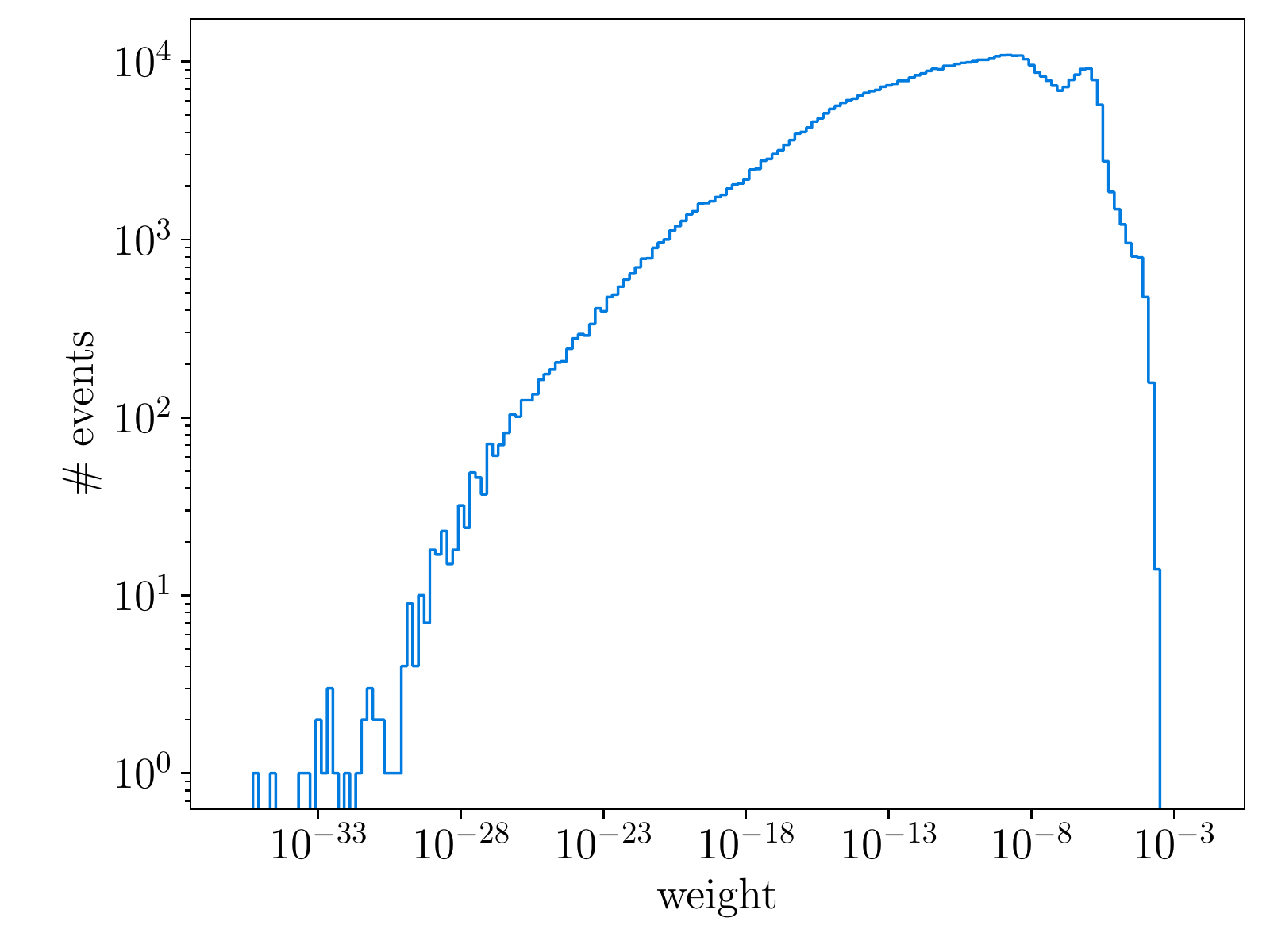}
  \includegraphics[page=1, width=.49\textwidth]{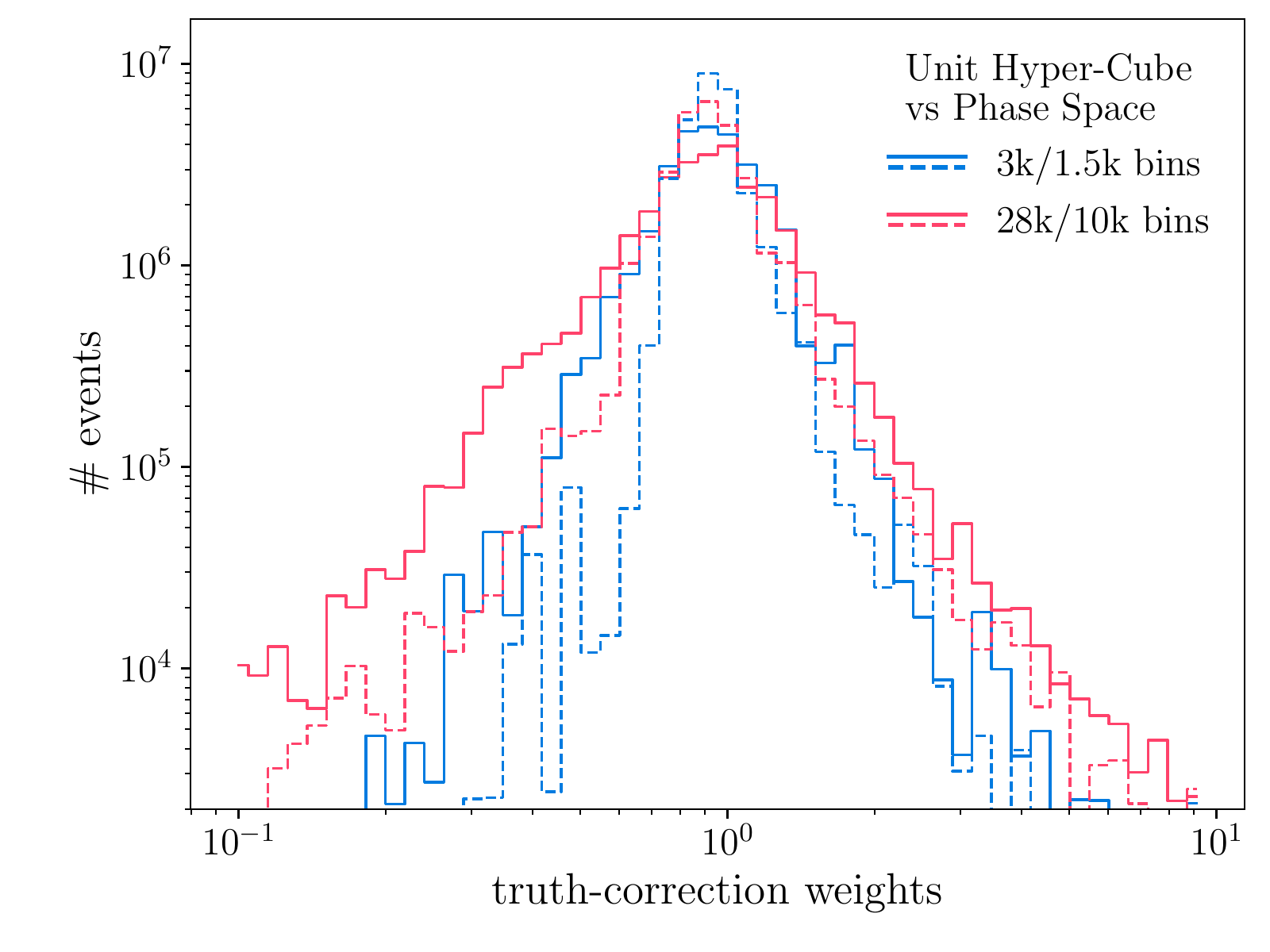}
  \caption{Left: Weight distribution for 500k weighted training
    events. Right: Truth-correction weights for 30M uwGAN events on
    the unit hyper-cube (solid) and the phase space parametrization of
    Eq.\eqref{eq:event-param} (dashed) for 2k (blue) and 14k (red)
    bins in 4-dimensions.}
  \label{fig:uw_drellyan_weights}
\end{figure}

\begin{figure}[t]
  \includegraphics[page=2, width=.49\textwidth]{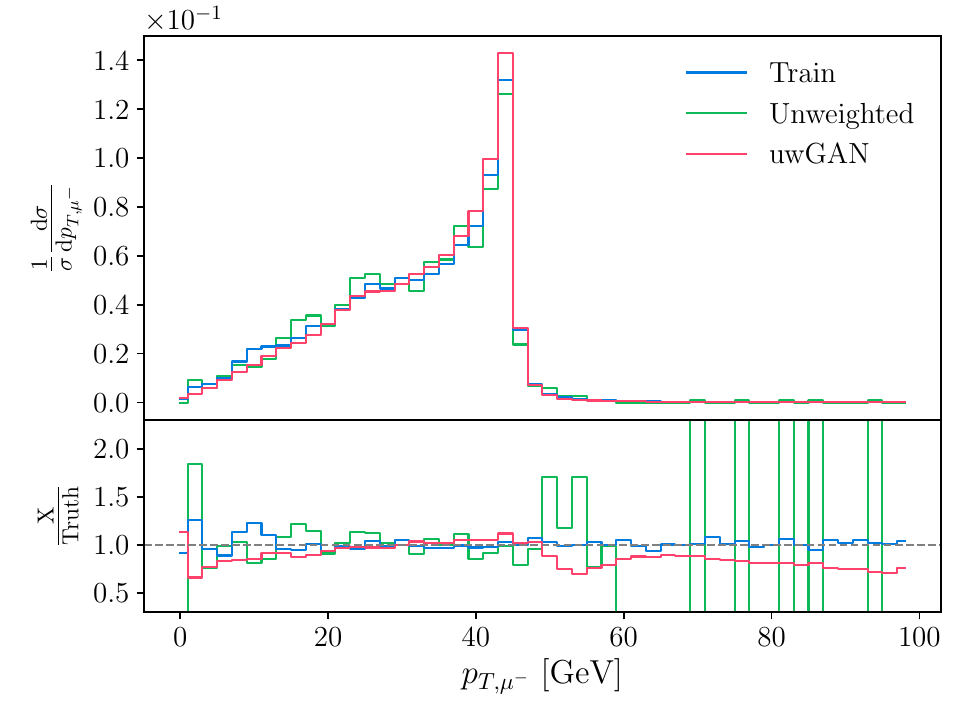}
  \includegraphics[page=7, width=.49\textwidth]{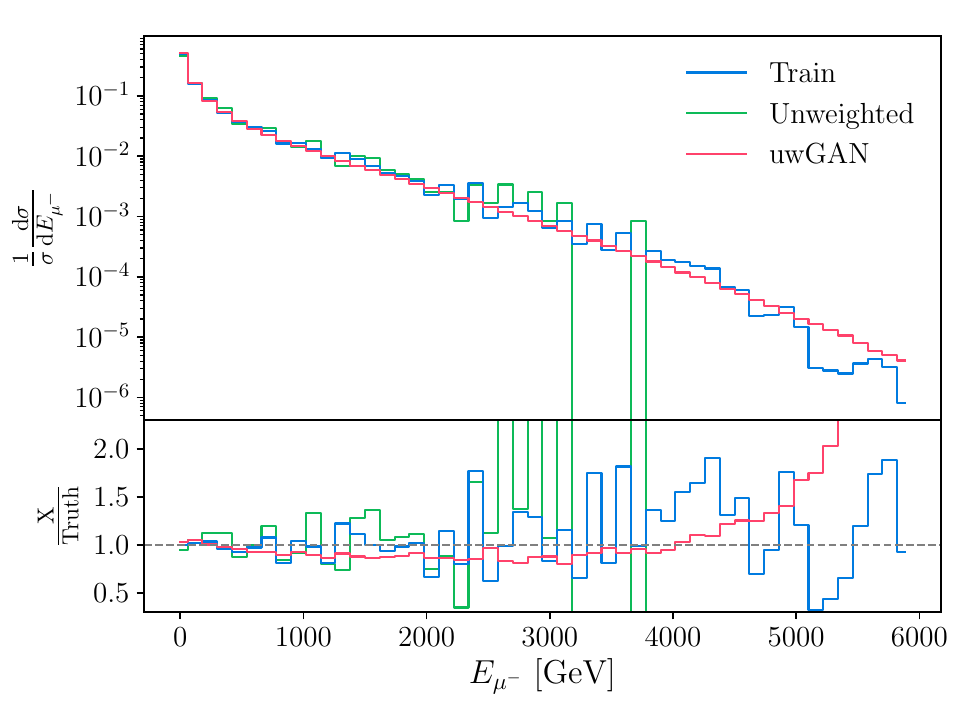} \\
  \includegraphics[page=4, width=.49\textwidth]{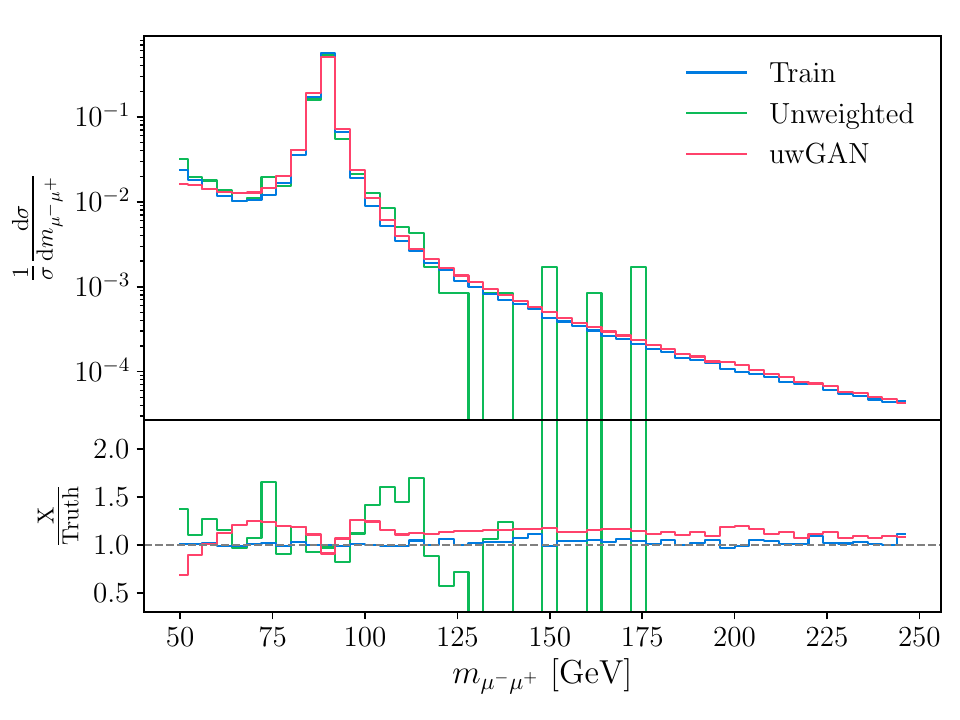}
  \includegraphics[page=6, width=.49\textwidth]{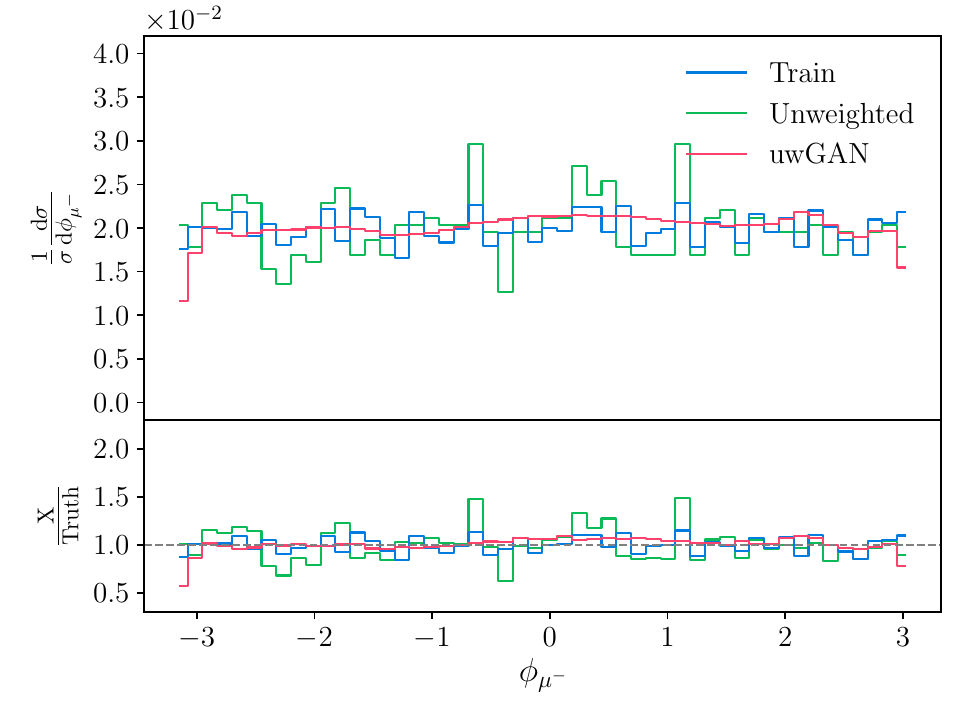}
  \caption{Exemplary kinematic distributions for the
    Drell-Yan process. For the kinematic distributions we show the
    500k weighted training events, 1k unweighted events using the
    standard unweighting algorithm discussed in
    Sec.~\ref{sec:unweight_basics}, and 30M uwGAN events.}
  \label{fig:uw_drellyan}
\end{figure}

In Fig.~\ref{fig:uw_drellyan_weights} we show the weight distribution
for our event generator, where the shown 500k event weights are
computed as the product of scattering amplitude, parton density, and
phasespace factor. While the distribution is very smooth, indicating
that the phase space is sampled precisely, the range of weights poses
a problem for an efficient event unweighting.  Even if we are willing
to ignore more than 0.1\% of the generated events, we still need to
deal with event weights from $10^{-30}$ to $10^{-4}$. Effects
contributing to this vast range are the $Z$-peak, the strongly
dropping $p_T$-distributions, and our deliberately poor phase space
mapping. The classic unweighting efficiency defined by
Eq.\eqref{eq:uw_efficiency} is 0.22\%, which is considered high for
state-of-the-art tools applied to complex LHC processes.  In the
following panels of Fig.~\ref{fig:uw_drellyan} we show a set of
kinematic distributions, first for the 500k weighted training events
including the deviation from a high-precision truth sample. Indeed,
this training data-set describes $E_\mu$ all the way to 6~TeV and
$m_{\mu \mu}$ beyond 250~GeV with deviations below 5\%. The perfectly
flat $\phi_\mu$ distribution turns out to be the challenge in our
specific phase space parametrization, with bin-wise deviations of up
to 20\% from the true distribution.

\begin{table}[b!]
	\begin{small} \begin{center}
			\begin{tabular}{l r}
				\toprule
				Parameter              & Value  \\
				\midrule
				Layers & 6\\
				Kernel initializer & He uniform\\
				G units per layer  & 414\\
				D units per layer & 187 \\
				G activation function & ReLU\\
				D activation function & leaky ReLU\\
				D updates per G & 2\\
				$\lambda_\text{wMMD}$ & $2.37$\\
				Learning rate & $0.0074$\\
				Decay & $0.42$\\
				Batch size & 1265 \\
				\midrule
				Epochs & 500 \\
				Iterations per epoch & 200\\				
				\bottomrule
			\end{tabular}
	\end{center} \end{small}
	\caption{Details for our uwGAN setup for the Drell-Yan process.}
	\label{tab:details_dy}
\end{table}

In addition to the unweighted training data, we also show the
kinematic distributions for unweighted events from a standard
algorithm. We use the hit-and-miss method described in
Sec.~\ref{sec:unweight_basics} without any further improvements, which
limits the number of unweighted events to 1000. Correspondingly, the
standard unweighted events only cover $E_\mu$ to 1~TeV and
$m_{\mu\mu}$ to 110~GeV. For $\phi_\mu$ the deviations also exceed
those of the training data significantly. This poor behavior is simply
an effect of the low unweighting efficiency and a serious challenge
for LHC precision simulations.

Alternatively, we can employ our uwGAN to unweight the Drell-Yan
training data. To take into account symmetries, we only generate
the degrees of freedom of the process. By construction, this guarantees
momentum conservation and on-shell conditions.  Before passing to the
discriminator both the generated batches $\{x_G\}$ and the truth
batches $\{x_T\}$ are parameterized as
\begin{align}
x=\{p_T ,p_{z_1}, p_{z_2}, \phi, w\},
\end{align}
where $w$ is the associated event weight. In order to reproduce the
sharp resonance appearing in the $m_{\mu\mu}$ distribution which
originates from the $Z$ boson propagator, we employ an additional
maximum mean discrepancy (MMD) loss~\cite{Butter:2019cae,mmd}. This
loss ensures that the network learns a pre-defined low-dimensional
function over the high-dimensional phase space. For the unweighting
GAN we generalize it in analogy to Eq.\eqref{eq:uwGAN},
\begin{align}
L_\text{wMMD}
&= \left[
\frac{\Langle  w(x)\,w(x')\,k(x, x') \Rangle_{x, x' \sim Q_T}}{\Langle  w(x)\,w(x') \Rangle_{x, x' \sim Q_T}}
+ \Langle  k(y, y') \Rangle_{y, y' \sim P_G}
-2 \frac{\Langle  w(x)\,k(x, y) \Rangle_{x \sim Q_T, y  \sim P_G}}{\Langle  w(x) \Rangle_{x \sim Q_T}} \right]^{\frac{1}{2}} \; ,
\end{align}
where we already use that $w_G(y)=1$. Note that we use MMD instead of
MMD$^2$ as this increases the sensitivity of the loss close to
zero. This loss is then added to the generator objective
\begin{align}
L_G\to L_G + \lambda_\text{wMMD}\,L_\text{wMMD}\;.
\end{align}
The network parameters are given in Tab.~\ref{tab:details_dy}. The
parameters in the upper panel have been determined by a random
hyperparameter search and have shown the best results.

In the right panel of Fig.~\ref{fig:uw_drellyan_weights} we again show
the truth-correction weights for our uwGAN events, evaluated on the
binned phase space either in terms of the unit hyper-cube ($r_j =
1~...~0$) or the appropriately cut phase space of
Eq.\eqref{eq:event-param}. The number of bins ignores empty bins and
shows the limitations of our bin-wise extraction of the truth
correction. While some of the truth corrections are not
negligible, we also know that they appear in the tails of the
generated phase space distribution and can easily be traced.  Even if
we consider the finite and bin-wise-defined truth correction with a
grain of numerical salt, we find the performance of our relatively
slim network quite convincing, given that we start from weighted
events with more than 25 orders of magnitude in weights. Most
importantly, the tails of the truth correction are a result of the
uwGAN unweighting, not a limiting factor like for the standard
unweighting procedure.

The appropriate measure of success for our uwGAN are the predicted
kinematic distributions. In Fig.~\ref{fig:uw_drellyan} we compare the
weighted training data, a corresponding unweighted event sample using
the standard algorithm, and the uwGAN results. In the lower panels we
show the relative differences to the truth, defined as a
high-statistics version of the training sample. While the training
data agrees with the truth very well, we see its statistical
limitations in the tail of the $E_\mu$-distribution. In addition, the
$\phi_\mu$ distribution for the weighted training data is 
noisier than one would expect for a smooth phase space.

In accordance with the established performance of GANs, the uwGANned
events reproduce the truth information well.  As always, the GAN
learns the phase space information only to the point where it lacks
training statistics and the GAN undershoots the true
distribution~\cite{Butter:2019cae}. This limitation can be
quantitatively improved by using different network
architectures~\cite{Bellagente:2020piv}. In our case it affects the
phase space coverage for $E_\mu \gtrsim 4.5$~TeV and $m_{\mu\mu}
\gtrsim 250$~GeV. These values are not quite on par with the training
data, but much better than for standard unweighting. For the same two
distributions we clearly see the loss of information from standard
hit-and-miss unweighting, and how the uwGAN avoid these large
losses. Along the same line, the $\phi_\mu$ distribution shows how the
uwGAN even slightly smoothes out the noisy training data.  We can
translate the reduced loss of information into a corresponding size of
a hypothetical hit-and-miss training sample, for instance in terms of
rate and required event numbers, and find up to a factor 100 for our
simple example.

\section{Outlook}
\label{sec:outlook}

First-principle precision simulations are a defining aspect of LHC
physics and one of the main challenges in preparing for the upcoming
LHC runs. Given the expected experimental uncertainties, we need to
improve both, the precision and the speed of the theory-driven event
generation, significantly to avoid theory becoming the limiting factor
for the majority of LHC analyses. One promising avenue is modern
machine learning concepts applied to LHC event generation.

In this study we proposed a significant improvement to one of the
numerical bottlenecks in LHC event generation, the unweighting
procedure. Such an unweighting step is part of every event generator,
and for complex final state it rapidly becomes a limiting factor. We
showed how to train a generative network on weighted events, with a
loss function designed to generate events of unit weights, or
unweighted events.

For a 1-dimensional and a 2-dimensional toy model we have shown that
our uwGAN can indeed be used for event unweighting and that in the
limit of perfect training it reproduces the true phase space
distributions just like standard methods like \vegas. While we cannot
beat the \vegas performance for a 1-dimensional test case, our uwGAN
easily circumvents \vegas limitations from the assumed dimensional
factorization.

As an LHC benchmark we use $\mu^+ \mu^-$ production and a poor
in-house event generator with a low unweighting efficiency over phase
space.  The uwGAN performs significantly better than the standard
unweighting procedure, both, in kinematic tails and for noisy training
data. Based on the success of our GAN architecture for top pair
production~\cite{Butter:2019cae} we expect our unweighting GAN to also
work for higher final-state multiplicities. While it is not clear how
much the speed gain from using an NN-unweighting in standard event
generators will be, this application of generative networks could be
easily implemented in the established LHC event generation chain. A
challenge for any phase-space-related network are processes with a
variable, llarge number of final-state particles, like $V$+jets
production~\cite{Bothmann:2020ywa,Gao:2020zvv}.  While we were
finalizing this study, similarly promising ideas were presented in
Ref.~\cite{Verheyen:2020bjw}, showing how generative networks benefit
from training on weighted events.

\begin{center} \textbf{Acknowledgements} \end{center}

We want to thank Aishik Gosh and David Rousseau for inspiring this
project.  RW acknowledges support by the IMPRS-PTFS and by HeiKA.  The
research of AB and TP is supported by the Deutsche
Forschungsgemeinschaft (DFG, German Research Foundation) under grant
396021762 — TRR 257 \textsl{Particle Physics Phenomenology after the
  Higgs Discovery}.

\bibliography{literature}

\end{document}